\documentstyle[11pt,newpasp,twoside]{article}
\def\edcomment#1{\iffalse\marginpar{\raggedright\sl#1\/}\else\relax\fi}
\reversemarginpar

\begin{document}
\title{Eclipsing Binary Pulsars}
 \author{Paulo C. C. Freire}
\affil{Arecibo Observatory, HC 3 Box 53995, Arecibo, PR 00612, USA}

\begin{abstract}
The first eclipsing binary pulsar, PSR~B1957+20, was discovered in
1987. Since then, 13 other eclipsing low-mass binary pulsars have been
found, 12 of these are in globular clusters.
In this paper we list the known eclipsing binary pulsars and their
properties, with special attention to the eclipsing systems in 47 Tuc.
We find that there are two fundamentally different groups
of eclipsing binary pulsars; separated by their companion masses.
 The less massive systems ($M_c \, \sim \, 0.02 \, \rm M_{\odot}$) are
 a product of predictable stellar evolution in binary pulsars.
The systems with more massive companions ($M_c \, \sim \, 0.2 \, \rm
M_{\odot}$) were formed by exchange encounters in globular clusters,
and for that reason are exclusive to those environments. This class of
systems can be used to learn about the neutron star recycling fraction
in the globular clusters actively forming pulsars. We suggest that
most of these binary systems are undetectable at radio wavelengths.
\end{abstract}

\section{Introduction}

The first eclipsing binary pulsar, PSR~B1957+12, was discovered in
1987 (\cite{fst88}). Since its discovery, 13 other eclipsing binary
pulsars with low-mass companions have been found, these are listed in
Table \ref{tab:binaries}, with relevant references. Of these systems,
12 are to be found in globular clusters (GCs). To these we add a list of binary
systems with companion masses smaller than 0.02~M$_{\odot}$ that are,
as we will see, intimately related to a subset of the eclipsing binary
systems.

The systems that have been known longest are also the ones for which
the pulsar has the largest flux density, it is therefore
understandable that such systems have been discussed with great
detail in a large number of publications. A review of all these
results is beyond the scope of this work, mainly because of
limitations in available space. We will concentrate
instead on the general properties of the population of eclipsing
binaries and infer some general trends. In section \ref{sec:47tuc},
we do concentrate specifically on the eclipsing binary pulsars in
47~Tuc; this is done for the sake of completeness, as most of the
remaining eclipsing binary systems are discussed elsewhere in these
proceedings.

In Table \ref{tab:binaries}, we included mostly references that
concentrate on the timing of the eclipsing pulsars. From these references,
one concludes that for the 5 eclipsing binaries with a published
timing baseline longer than 4 years (PSR~B1957+20 and
PSR~J2051$-$0827, Terzan 5 A and 47~Tuc~J and O) random variations of
the orbital period have invariably been detected.

\begin{table}
\caption{\label{tab:binaries}
A set of eclipsing binary pulsars and related objects in GCs and in
the Galactic disk. Those
for which a timing solution is not known are followed by an asterisk,
the names could change in the future when that solution is
published. 47~Tuc~W has a position determined by optical
astrometry. Those that are known to eclipse are followed by letter
``e''. The minimum companion masses ($m_c$) are calculated
assuming a pulsar mass of 1.4 M$_{\odot}$. References are:
rh: Ransom, Hessels, these proceedings,
dn: David Nice, these proceedings, st: Stappers, these proceedings,
po: Possenti, these proceedings
1:~\cite{clf+00}, 2:~\cite{pdm+03}, 3:~\cite{lbhb93},
4:~\cite{dpm+01,fpds01}, 5:~\cite{lmbm00,nat00}, 6:~\cite{rsb+04},
7:~\cite{fck+03}, 8:~\cite{cha03}, 9:~\cite{dlm+01,rgh+01},
10:~\cite{dma+93}, 11:~\cite{fst88,arz95}, 12:~\cite{sbl+96,sbm+98}
}
\begin{tabular}{ l l c c c l }
 & & & & & \\
\hline
\hline
Pulsar & GC, letter & P(ms) & $P_b$ (days) & $m_c (M_{\odot})$ & Ref. \\
\hline
J0024$-$7204V* & 47~Tuc~V (e?) & 4.81  & $\sim$0.2 & $\sim$0.3 & 1 \\
J0024$-$7204W* & 47~Tuc~W (e)  & 2.35  & 0.133     & 0.127 & 1 \\
J1701$-$3006B & M62~B (e)      & 3.59  & 0.145     & 0.124 & 2 \\
B1718$-$19    & NGC~6342A (e)  & 1004  & 0.258     & 0.117 & 3 \\
J1740$-$5340  & NGC~6397A (e)  & 3.65  & 1.354     & 0.188 & 4,po \\
J1748$-$2446A & Terzan 5A (e)  & 11.56 & 0.076     & 0.089 & 5,dn \\
J2140$-$2310A & M30~A (e)      & 11.02 & 0.174     & 0.101 & 6,rh \\
\hline
J0024$-$7204I & 47~Tuc~I       & 3.48  & 0.230     & 0.013 & 7 \\
J0023$-$7203J & 47~Tuc~J (e)   & 2.10  & 0.121     & 0.021 & 7 \\
J0024$-$7204O & 47~Tuc~O (e)   & 2.64  & 0.136     & 0.022 & 7 \\
J0024$-$7204P* & 47~Tuc~P      & 3.64  & 0.147     & 0.018 & 1 \\
J0024$-$7204R* & 47~Tuc~R (e)  & 3.48  & 0.066     & 0.026 & 1 \\
J1518+0204C*   & M5~C (e)      & 2.48  & 0.087     & 0.038 & rh \\
J1701$-$3006E* & M62~E (e)     & 3.23  & 0.16      & 0.03  & 8 \\
J1701$-$3006F* & M62~F         & 2.30  & 0.20      & 0.02  & 8 \\
J1807$-$2459A* & NGC~6544~A    & 3.06  & 0.071     & 0.009 & 9 \\
B1908+00*      & NGC~6760~A    & 3.62  & 0.141     & 0.018 & 10,dn \\
J1953+1846A*   & M71~A (e)     & 4.89  & 0.177     & 0.032 & rh \\
B1957+20       & (Galaxy) (e)  & 1.61  & 0.382     & 0.022 & 11,dn \\
J2051$-$0827   & (Galaxy) (e)  & 4.51  & 0.099     & 0.027 & 12,st \\
\hline
\hline
\end{tabular}
\end{table}

\section{Four eclipsing binary systems in 47~Tuc}
\label{sec:47tuc}

Of the total of 22 millisecond pulsars (MSPs) in 47 Tuc observed at
Parkes, 15 are members of
binary systems. Of these, at least 4 are eclipsing: 47~Tuc~J, O, R, W
and possibly V. There are two other systems with very short orbital
periods and very small companion masses that are analogous to
47~Tuc~J, except for their even smaller mass functions and the
non-detection of eclipses; these are 47~Tuc~I and P. The systems with
known timing solutions (47~Tuc~I, J and O) are analyzed in detail in
Freire~et~al.~(2003)\nocite{fck+03}. The remaining systems (47~Tuc~P,
R, V and W) are briefly described in Camilo et al. (2000)\nocite{clf+00}.

\subsection{47~Tuc~J}

This system has only been confirmed to eclipse at 430~MHz,
at this frequency the pulsar is
never detected near superior conjunction. The pulsar
is detectable in some 660-MHz observations near superior conjunction,
in others it is not. This could be due to occasional eclipses, but it
could also be due to scintillation. At 1400~MHz no eclipses are ever
detectable. In any case, it is clear that the
DM of the pulsar varies with orbital phase, being measurably higher
at superior conjunction.
\begin{figure}
\setlength{\unitlength}{1in}
\begin{picture}(0,3.4)
\put(1.1,0.0){\includegraphics{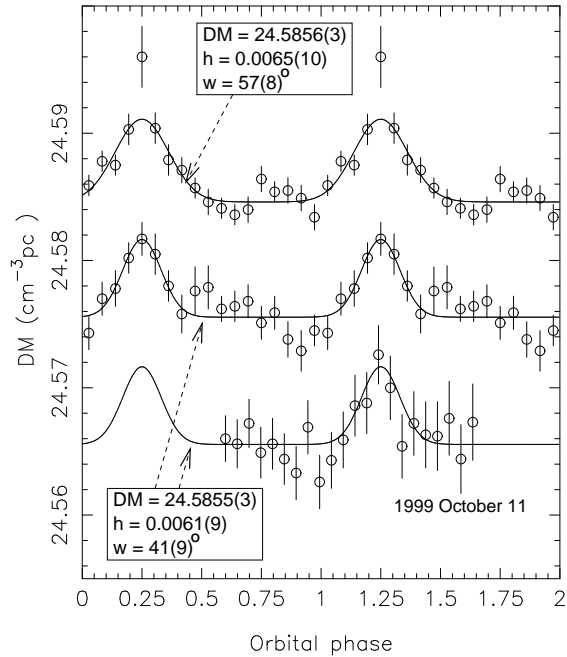}}
\end{picture}
\caption{\label{fig:dJ} DM as a function of orbital phase ($\phi = 0$
is the ascending node) for 47~Tuc~J. The upper plot represents the
values for DM obtained from 1998 January data at 660\,MHz and 1400-MHz
data obtained from 1998 June to 1999 August. The middle plot represents
the measurements made with 1999--2002 high-resolution 1400-MHz data,
displayed 0.01~cm$^{-3}$pc below their measured values for
clarity.  The third plot represents the DMs obtained using only the
data from the best observation (1999 October 11), displayed
0.02~cm$^{-3}$pc below their measured values. The numbers indicate the
parameters of the best fit for each data set (see text), which is
depicted by the solid curves. From Freire et al. (2003).}
\end{figure}
The extra electron column density at superior conjunction is $\sim 1.7
\times 10^{16}\rm\,cm^{-2}$, which is about 10 times smaller than the
extra electron column density observed for PSR~J2051$-$0827 at the
same orbital phase (\cite{sbl+01}).

Considering the separation between the pulsar and
its companion for inclinations near 90$^\circ$, $a = 1.14\,R_{\odot}$,
and the length of the eclipse at 430\,MHz, the radius of the eclipsing
object must be larger than $0.7\,R_{\odot}$.  This implies an average
electron density of $\sim 10^5\,$cm$^{-3}$ near the companion.
The Roche lobe for the companion is given by (\cite{egg83}):
\begin{equation}
R_L = \frac{0.49 \times a q^{2/3}}{0.6 q^{2/3}+ \ln ( 1 + q^{1/3} ) }
\end{equation}
where $q\,=\,m_c/m_p$ and $a$ is the separation between the pulsar and
the companion. The only source of uncertainty is the inclination of
the system. For $i\,=\,90^\circ$, $m_c\,=\,0.0209\,\rm M_{\odot}$,
$q\,=\,0.0155$, $a\,=\,1.14\,\rm R_{\odot}$ and
$R_L\,=\,0.13\,R_{\odot}$. For $i \,=\,60^\circ$, $m_c\,=\,0.0241\,\rm
M_{\odot}$, $q\,=\,0.0179$, $a\,=\,1.14\,\rm R_{\odot}$ and
$R_L\,=\,0.14\,R_{\odot}$. Therefore, the Roche Lobe is much smaller
than the diameter of the plasma cloud; the matter
responsible for the increased DMs is not bound to the companion
object. This is a clear indication that the companion is loosing mass,
it is behaving like a comet. The same is true for the other
eclipsing systems discussed in this text.

For 47~Tuc~J, we can measure
significant orbital evolution: $\dot{P_b} = (-0.52 \pm 0.13) \times 10^{-12}$
and $\dot{x} = (-2.7 \pm 0.7) \times 10^{-14}$. The $\dot{P_b}$ is
much smaller than what was measured for the two Galactic VLMBPs, but
this could be due to the small timescale of the observations made to date.

\subsection{47~Tuc~O}

\begin{figure}
\setlength{\unitlength}{1in}
\begin{picture}(0,8.5)
\put(1.0,0.0){\includegraphics{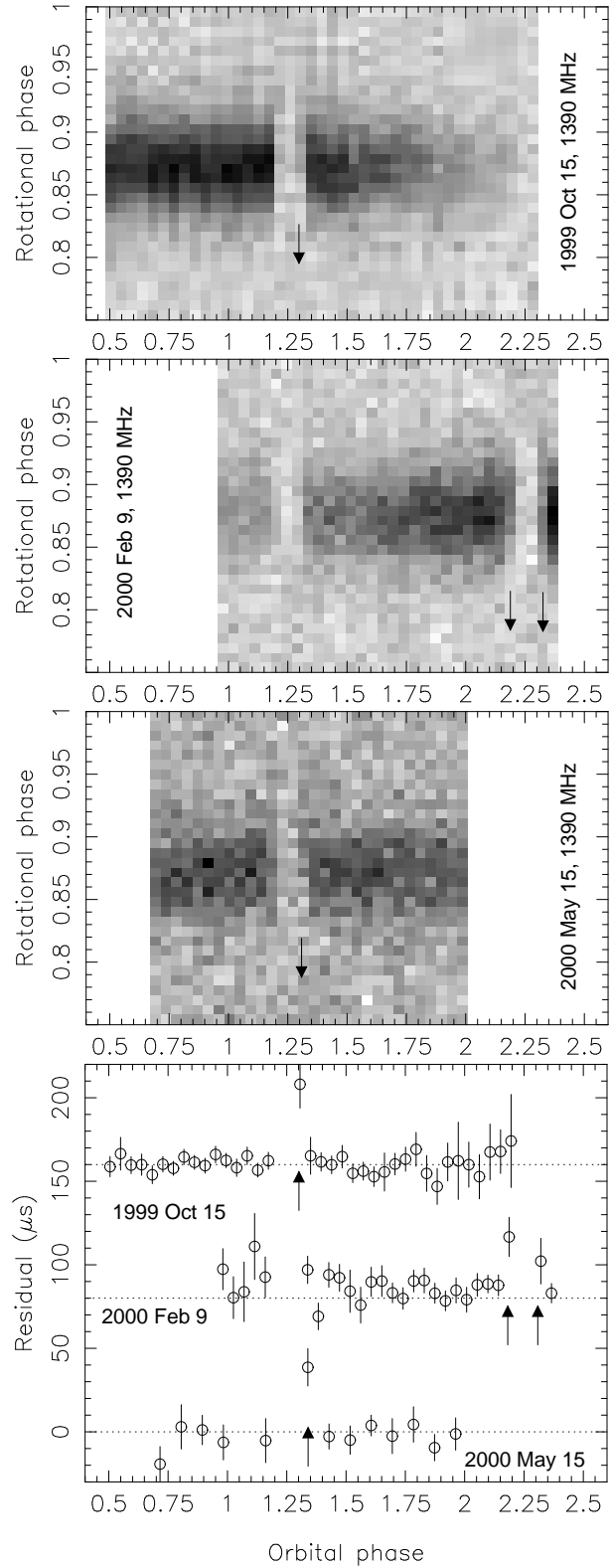}}
\end{picture}
\caption{\label{fig:fO} Top plots: Intensity as a function of orbital
and rotational phases for three of the best observations of 47~Tuc~O.
The darkness is linearly proportional to the measured intensity. The
residuals for the corresponding TOAs (used to obtain the best fit) are
displayed at bottom. Bottom plot: residuals for the same
observations. From Freire et al. (2003).}
\end{figure}

This pulsar is different from 47~Tuc~J in the sense that it always
displays sharp, well-defined eclipses at 1400~MHz (see
Fig. \ref{fig:fO}). There are no good detections of this pulsar at
660~MHz, the reason for this lack of detections is unknown, the most
likely explanation is that this pulsar has an unusually flat spectral
index. The pulsar exhibits strong variability of its orbital
parameters: $\dot{P_b} = (9 \pm 1) \times 10^{-12}$, $\ddot{P_b} = (24
\pm 8) \times 10^{-20}\,$s$^{-1}$, and $\left| \dot{x} \right| < 1.8
\times 10^{-13}$. This is comparable in timescale and amplitude with the
variations observed for the Galactic eclipsing pulsars, PSR~B1957+20
and PSR~J2051$-$0827. It can therefore be said that 47~Tuc~J and O are
similar to the two eclipsing binaries in all their observational
characteristics, which suggests a common origin to these systems.

\begin{figure}
\setlength{\unitlength}{1in}
\begin{picture}(0,3.5)
\put(0.2,3.4){\includegraphics{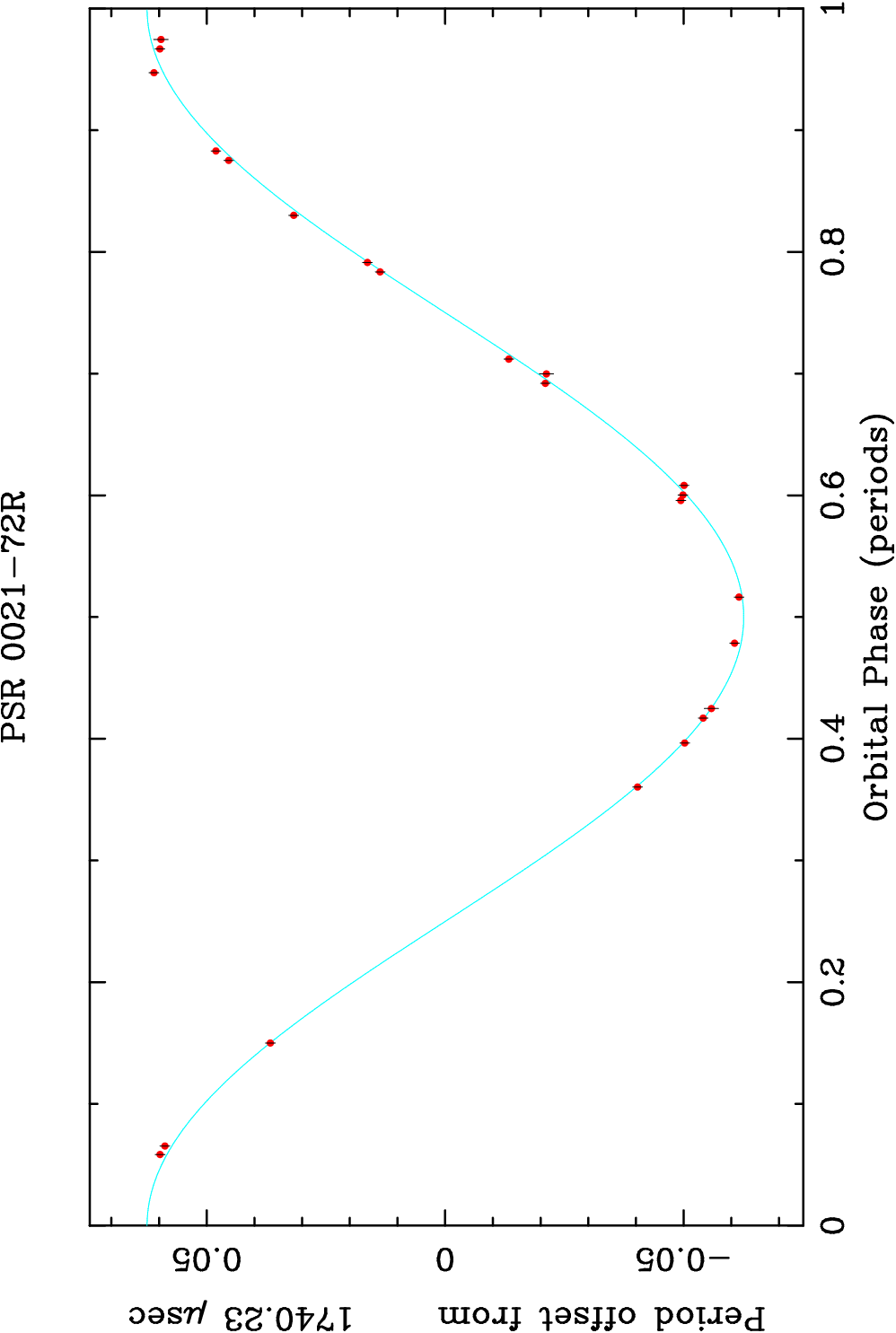}}
\end{picture}
\caption{\label{fig:R} Orbital model for 47~Tuc~R. The pulsar has now
  been detected on 5 different orbits at 1400-MHz, the eclipses never
  fail to happen.}
\begin{picture}(0,3.5)
\put(0.2,3.4){\includegraphics{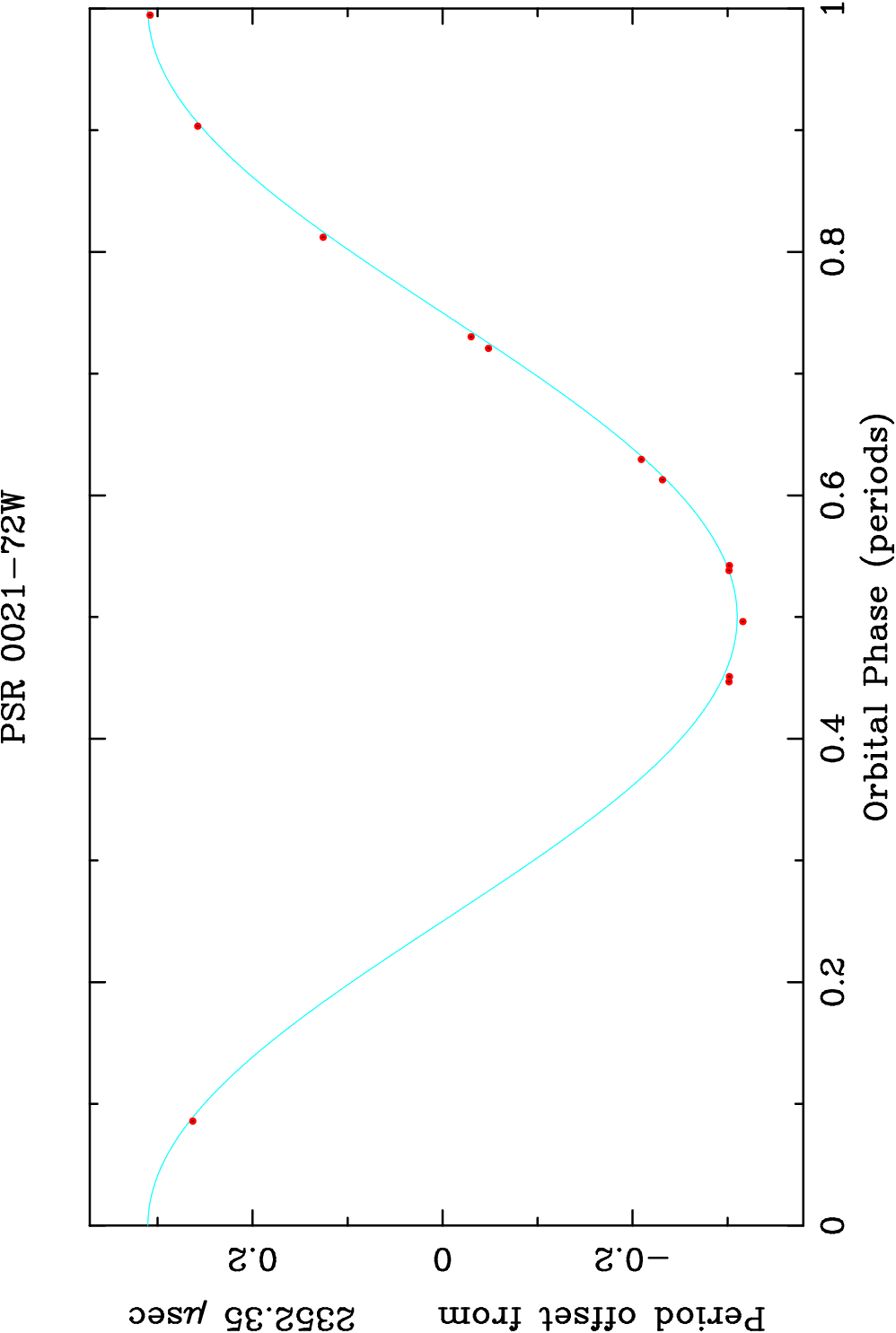}}
\end{picture}
\caption{\label{fig:W} Orbital model for 47~Tuc~W. The pulsar has now
  been detected on 4 different orbits at 1400~MHz. The eclipses last
  for about 40\% of the whole orbit, and never fail to happen.}
\end{figure}

\subsection{47~Tuc~R}

This is still the known binary pulsar with the shortest orbital
period (96 minutes), the minimum companion mass is about
0.03~M$_{\odot}$. When first announced in Camilo~et~al.~(2000), only
one orbit had been detected; in that orbit the pulsar signal was
absent at superior conjunction in a way that is very characteristic of
an eclipsing binary. Since then, the pulsar has been regularly
detected in the high-resolution 20~cm data, which has allowed us to
obtain orbital phase connection (see Figure \ref{fig:R}). The
pulsar has eclipsed near superior conjunction on all observations where
it is detectable.

\subsection{47~Tuc~W}

This binary pulsar has an orbital period of 3.19 hours. Assuming for
the pulsar a mass of 1.35~M$_{\odot}$, the companion has a minimum
mass of $\sim$0.13~M$_{\odot}$, which is unusually large when compared
to the previously known eclipsing binaries in 47~Tuc. At least in this
respect this object is similar to the previously known Terzan~5~A
binary system. The two radio eclipses in the single detection reported
by Camilo et al. (2000) last for
35-40\% of the orbital period. The companion has since been detected
at optical wavelengths with the HST based on the coincidence of
orbital periods (\cite{egc+02}); it is a normal main-sequence (MS)
star. Since then the pulsar has been
detected several times in the high-resolution Parkes 20~cm data (see
Figure \ref{fig:W}). The pulsar has eclipsed near superior
conjunction on all observations when it is detectable.

\subsection{Short-period binaries with small-mass companions}

For 47~Tuc~I, the pulsar with the (apparently) lightest
companion in 47~Tuc, the 1400-MHz TOAs are well described by a
circular orbit, no eclipses are observed for superior
conjunction. The number and quality of the TOAs at superior
conjunction is the same as at other orbital phases.
47~Tuc~P is similar in many respects to PSR~B1908+00 in NGC~6760
(\cite{dma+93}, see also David Nice's paper, these proceedings);
during their detections (all at 1400\,MHz, of which there is only one
for 47~Tuc~P) none of these pulsars displayed eclipses.

\section{Discussion}
\label{sec:discussion}

\begin{figure*}
\setlength{\unitlength}{1in}
\begin{picture}(0,5.9)
\put(0.2,0.0){\includegraphics{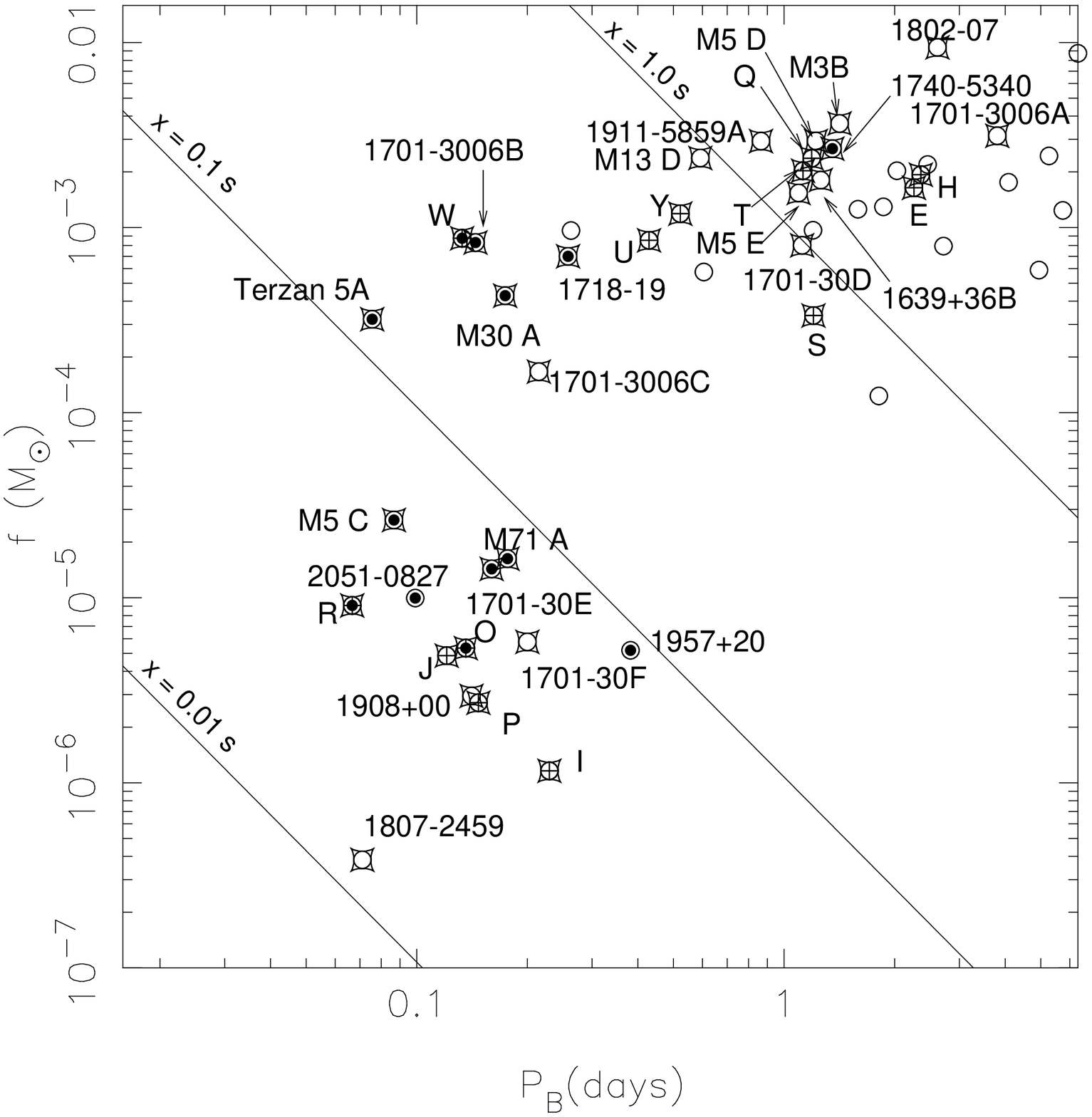}}
\end{picture}
\caption{\label{fig:fPB} Mass function plotted against orbital period
for all the binary pulsars known (circles) within the given range of
orbital period and mass function. Those binaries in GCs
are indicated inside 4-pointed stars and named; we highlight those in
47~Tuc with a ``+'' and name them only by their letters. A black dot
inside the symbol indicates a binary pulsar that eclipses at 1400 MHz;
the two eclipsing Galactic binaries are also named. Binaries with
$f\,<\,3 \,\times\,10^{-5}\,$M$_{\odot}$
(VLMBPs) have companions with masses of
0.01--0.04 M$_{\odot}$, and generally present shorter orbital periods
than the LMBPs ($f\,>\,10^{-4}\,$M$_{\odot}$). About two thirds
of the VLMBPs display eclipses. The inclined
lines indicate constant projected semi-major axis $x$. }
\end{figure*}

In Figure \ref{fig:fPB}, we can see that these eclipsing binary
systems are divided in two main categories. The first set, consisting
of a total of 8 objects, has mass functions below
3$\,\times\,10^{-5}\,\rm M_{\odot}$. These belong to a class of
objects that we will henceforth call very low-mass binary pulsars
(VLMBPs), also known as ``black-widow'' binaries. We will use the
VLMBP designation regardless of the presence of eclipses.
The second class of eclipsing binaries has mass functions above
3$\,\times\,10^{-4}\,\rm M_{\odot}$, they will be henceforth be called
eclipsing low-mass binary pulsars (ELMBPs). A third class of eclipsing
binary pulsars has high-mass MS companions, they have a
single known representative, PSR~B1259$-$63. This system is discussed
by Simon Johnston in these proceedings.

\subsection{Very low-mass binary pulsars}

\subsubsection{Formation}

The clear separation in terms of mass function between the LMBPs
and the VLMBPs strongly suggests that these binaries form a population
that is distinct in its evolutionary history. This is expected from
simulations of binary evolution in the GC 47~Tuc (\cite{rpr00}), which
have suggested that these objects were once normal MSP-WD binaries
that underwent further recycling because of orbital decay due to the
emission of gravitational waves.
Such a scenario is consistent with the observed rotational periods of
these objects. PSR~B1957+20 has a rotational period of 1.6 ms, the second
shortest known. The recently discovered PSR~J1953+1846A, in M71, has
the longest rotational period of this class, at 4.9 ms. Most of the
remaining rotational periods range between 2 and 3 ms.
It is therefore fair to say that, as a class, the pulsars in VLMBPs
are the most heavily recycled neutron stars known, which supports the
idea of more than one prolonged and intense accretion episode.

This formation process should also be active in the Galaxy, as it
results from normal stellar evolution. The fact that we find two
objects in the disk of the Galaxy (PSR~B1957+20 and PSR~J2051$-$0827)
similar in all their distinguishing features (eclipses and orbital
variability) to the VLMBPs in clusters supports this idea. It is
likely, however, that exchange encounters in GCs were involved in the
formation of the progenitor binary systems, this is is necessary to
explain the large numbers of these systems in GCs.

\subsubsection{Non-eclipsing VLMBPs}

The VLMBPs with the lowest mass functions do not display eclipses,
despite having orbital periods similar to those of the eclipsing VLMBPs. 
A correlation between small mass function and the lack of eclipses can
only be understood if the mass distribution is narrower than what is
indicated in Table \ref{tab:binaries} and the
inclination is the main factor determining the
observed mass function for these objects: at high inclinations, the
system is seen edge-on, eclipses can be seen and the minimum companion
mass is near 0.03 to 0.04~M$_{\odot}$. At lower inclinations, no eclipses can
be seen, as we are looking at the system nearly face-on; the mass function
is lower by a $(\sin i)^3$ factor. Being the same sort of object,
VLMBPs with no eclipses should exhibit the same kind of orbital
variability as those with eclipses. Table \ref{tab:binaries} shows
that, with the exception of 47~Tuc~I, none of these systems has
published timing solutions. The timing precision for 47~Tuc~I is quite
low, even it if had the same amount of variability seen in the other
systems, the variations of orbital period would not be detectable.

It is possible that the above interpretation of the role of the
orbital inclination is incorrect and that
companions with a minimum mass of $\sim 0.01$\,M$_{\odot}$ are
fundamentally different in nature that those with $\sim
0.03$\,M$_{\odot}$, being for some reason incapable of producing any
sizable atmospheres. We find such an explanation unlikely, as the
less massive companions must have lower surface gravities and escape
velocities; it should be easier, using the same energy input from the
pulsar, to create an extended atmosphere around such objects.

\subsection{Eclipsing low-mass binary pulsars}

There are six known ELMBPs. Optical identifications of the companions
were made for three systems: PSR~J0024$-$7204W in 47~Tucanae
(\cite{egc+02}, Grindlay, these proceedings), PSR~B1718$-$19 in
NGC~6349 (\cite{kkk+00}) and and PSR~J1740$-$5340 in NGC 6397
(\cite{fpds01}, Possenti, these proceedings).
In all cases the companion happens to be
a low-mass MS star. It is therefore reasonable to conclude
that this is also the case for the three remaining systems,
PSR~J1748$-$2446A, in Terzan 5, PSR~J1701$-$3006B in M62 and
PSR~J2140$-$2310A, in M30. In fact, there is a tentative optical
identification of the companion of PSR~J2140$-$2310A with a faint, red
MS star (\cite{rsb+04}). Interestingly, none one of the
non-eclipsing low-mass binary pulsars (henceforth LMBPs) was found to
have a MS star as a companion, in all cases for which we
have optical identifications (e.g., PSR~J0024$-$7203U in 47~Tuc,
\cite{egh+01}, PSR~J1911$-$5859A, in NGC~6752, Cees Bassa et al.
these proceedings, and PSR~B1620$-$26, in M4, \cite{srh+03}, also in
these proceedings) the companion turns out to be a
low-mass white dwarf (WD), presumably the remnant of the star from which
the MSP has accreted and got recycled. LMBPs and ELMBPs are therefore
very different kinds of objects.

\subsubsection{Formation}

The nature of ELMBPs suggests that these objects were formed through
exchange encounters. These only have a reasonably probability of
happening in GCs, therefore ELMBPs should only occur in
GCs, as observed. This is not the case with LMBPs and
VLMBPs, which are also observed in the Galaxy and are therefore the
result of normal stellar evolution in binary systems.

In an exchange encounter, an isolated star or a component of a 
binary system comes within less than 4 times the separation between
the components of a binary system, this is normally followed by
strong, unpredictable gravitational interactions. The end result tends
to be the formation of a new binary
system containing the two most massive objects, and the ejection of
the lighter star(s), with a resulting recoil of the newly formed binary
system. The tighter the system, the more violent this recoil will be.

Originally, GCs had many massive, blue MS stars. These
reached the end of their lives very early in the cluster life, with a
large number of supernova explosions. These formed many energetic,
young pulsars, these slowed down and died on a timescale much shorter
than the age of the cluster. Many remained bound to the cluster, and
roam it today as dead neutron stars. Occasionally, exchange encounters
place one of these neutron stars in orbit with a MS star, then the
evolution of the MS star might lead it to fill its Roche lobe, leading
to accretion onto the neutron star. Such systems are observed in the
Galactic disk and in  GCs as low-mass X-ray binaries, but are far more
abundant per unit mass in GCs; in the Galactic disk these systems
can only form from primordial binary systems. After accretion stops,
these systems become MSP - WD binary pulsars (LMBP), which are also
over-abundant in GCs. They can then undergo further exchange encounters.

It happens sometimes in GCs that the neutron star placed in orbit
around a MS star through an exchange encounter is a previously
recycled pulsar, in some cases a former member of a LMBP
(channel I), in others a previously isolated pulsar that intruded into
a primordial MS binary system (channel II). These are the
systems we observe as ELMBPs. They are very similar to the systems
that give origin to LMXBs, but it is not clear that such will be their
fate: pulsar wind pressure might prevent accretion onto the neutron star.
For a system like PSR~B1718$-$19, with a mildly recycled pulsar with a
characteristic age of only a few million years, this will not be a
problem after pulsar emission ceases, which should happen in a
relatively small amount of time, such a system is therefore a prime
candidate to become an LMXB in less than a Gyr.

\subsubsection{Position relative to the center of the GCs}

It is interesting to remark that two of the ELMBPs are found to be at
considerable distances from the centers of their GCs. Such are
the cases of PSR~J1718$-$19 and PSR~J1740$-$5340. This is suggestive
of exchange encounters, we are probably witnessing the
effects of stellar recoil. After evolving for a few more hundred Myr,
two-body encounters with other stars will gradually reduce the kinetic
energy of the system and bring it closer to the center of the GC,
as observed for the remaining 4 ELMBPs and most of the other pulsars
in GCs. For system like
PSR~J1718$-$19, the characteristic age of the pulsar is smaller
than the time it will take for the system to settle at the GC's
core, it is probable that when that happens this object will no longer
be observable as a radio pulsar. Conversely, it is unlikely that one
can observe a system like PSR~J1718$-$19 near the center of a
GC.

\subsubsection{Rotational period distribution}

While VLMBPs have a very narrow range of rotational periods, ELMBPs
have a very wide distribution of rotational periods. 47~Tuc~W has a
rotational period of 2.35 ms, while PSR~B1718$-$19 has a rotational
period of 1.004 seconds, the largest rotational period for as known
pulsar in a 
GC. The other rotational periods are 11.6, 11.0, 3.59 and 3.65 ms.
 This is, again, consistent with the idea of these systems forming
through exchange encounters; the present orbital companion and
orbital parameters have little relation to the recycling history of
the pulsar.

\subsubsection{The significance of 47 Tuc V}

A perplexing feature of Fig. \ref{fig:fPB} is
that, despite their fundamentally different nature, the ELMBPs have
the same mass functions as the LMBPs, implying in both cases companion
masses of 0.1 to 0.2~M$_{\odot}$. Such a uniform number is probably
not unimaginable for LMBPs, as the stellar evolution process they went
through is similar for all systems. However, if ELMBPs were formed in
exchange encounters, then the MS stars in their orbits should have the
same mass distribution as the MS stars in the parent cluster.

A possible hint to the solution of this problem is the existence of
47~Tuc~V, described in Camilo et al. (2000). This pulsar has a
rotational period of 4.81 ms. {\em Assuming} that the orbit is
circular, the orbital period is about 5 hours and the companion mass
is about 0.3~M$_{\odot}$. If this model is correct, the pulsar was
only observed near inferior conjunction. Even at this orbital phase,
the pulsar emission is sharply interrupted and re-started, with
timescales of about 5 minutes. This suggests a relatively large
MS companion, and the presence of orbiting clouds of
material that occult the pulsar even at inferior
conjunction. The mere existence of this system and its phenomenology
strongly suggest that i) Pulsars with MS companions
heavier that 0.3~M$_{\odot}$ exist and ii) such systems are likely,
because of the large size of the star, to be ``shrouded'' by material
from the star. Therefore, the lack of observations of systems with MS
companions more massive than 0.2~M$_{\odot}$ might at least in part be
due to an adverse selection effect.
This shrouding effect should not be so severe for pulsars in wider
binary systems. It should be possible to detect pulsars with, say,
0.7-M$_{\odot}$ MS companions if the orbital periods are of the order
of a few days, but a detection is probably unlikely if the orbital
period is of the order of a few hours.
It is possible that pulsars in very tight orbits with relatively
massive MS stars might in their future undergo a common envelope
phase, forming something similar to a LMBP or VLMBP.

\subsubsection{Fraction of recycled neutron stars in GCs}

Of a total of 80 known pulsars in GCs, the six known
ELMBPs are the result of exchange encounters that placed the pulsar in
orbit around a MS star; i.e., pulsars that became part of a
system that could evolve towards a LMXB and later into a LMBP
if the pulsar had been a dead neutron star to start with. We can think
of these 80 pulsars as pre-existing probes of neutron star
behavior. If 7.5\% of these pulsars suffered an exchange encounter
that puts them, for the second time, on a potential recycling path, we
can suggest that the same happened to the unseen dead neutron stars,
i.e., the {\em observable} pulsars in GCs (the total
number of active pulsars that one would find assuming no limitations in
sensitivity) are about 7.5\% of the total neutron star population in
GCs. This estimate does not take into account effects
that would increase the percentage of recycled pulsars compared to
this number:
\begin{itemize}
\item 7.5 \% is an under-estimate of the number of ELMBPs due to
selection effects against the detection of this kind of objects,
including the shrouding effect discussed in the previous paragraph.
\item Single, dead neutron stars have been available for exchange
interactions for a longer time than the recycled systems, which
necessarily formed later.
\end{itemize}
It does not take into account effects that would diminish this
 percentage either:
\begin{itemize}
\item Dead, neutron stars are generally isolated, so they
have no companion to exchange for an isolated
MS star. To find a companion, they depend
exclusively on the existence of primordial MS-MS binaries.
\end{itemize}
Making a better allowance for these selection
effects and extending the samples through deeper surveys should allow
us to derive a better fraction for neutron star recycling in 
GCs. However, such a number will always be to some extent 
cluster-dependent, as
some are denser and therefore bound to have a higher recycling
rate and produce more pulsars. The global figure of 7.5\% relates
mostly to the GCs actively forming pulsars.

\section{Conclusions}

We have listed the set of observational properties that lead us to
believe that VLMBPs, LMBPs and ELMBPs are different sorts of
objects. The first two classes are likely products of stellar
evolution in binary systems, and can therefore be observed in 
GCs and the Galaxy. The second kind is likely the result of
exchange encounters, such systems should only be
observed in GCs. Many of them are unlikely to be
undetectable, because of the effect of shrouding suggested by the
existence of 47~Tuc~V. The rate of occurrence of such systems in GCs
can be used to estimate neutron star recycling for the GCs that
produce more pulsars.

\end{document}